\documentclass[twocolumn]{aastex631}

\usepackage{graphicx}
\usepackage{dcolumn}
\usepackage{bm}
\usepackage{color}
\usepackage{ulem}
\definecolor{bazaar}{rgb}{0.6, 0.47, 0.48}
\definecolor{auburn}{rgb}{0.43, 0.21, 0.1}
\definecolor{antiquefuchia}{rgb}{0.57, 0.36, 0.51}
\definecolor{ao}{rgb}{0.0, 0.5, 0.0}
\definecolor{blush}{rgb}{0.87, 0.36, 0.51}
\definecolor{dp}{rgb}{0.5, 0, 0.5}

\shorttitle{Approach to equilibrium}
\shortauthors{Williams \& Hjorth}

\graphicspath{{./}{figures/}}

\begin{document}

\title{Statistical Mechanics of Collisionless Orbits. V. The approach to equilibrium for idealized self-gravitating systems}

\author{Liliya L. R. Williams}
\affiliation{School of Physics and Astronomy, University of Minnesota, 116 Church Street, Minneapolis, MN 55455, USA}

\author{Jens Hjorth}
\affiliation{DARK, Niels Bohr Institute, University of Copenhagen, Jagtvej 128, 2200 Copenhagen, Denmark}


\begin{abstract}
Self-gravitating Newtonian systems consisting of a very large number of particles have generally defied attempts to describe them using statistical mechanics. This is paradoxical since many astronomical systems, or simulations thereof, appear to have universal, equilibrium structures for which no physical basis exist. A decade ago we showed that extremizing the number of microstates with a given energy per unit mass, under the constraints of conserved total energy and mass, leads to the maximum entropy state, $n(E) \propto \exp (-\beta(E-\Phi_0))-1$, known as DARKexp. This differential energy distribution, and the resulting density structures, closely approximate those of dark-matter halos with central cusps, $\rho \sim r^{-1}$, and outer parts, $\rho \sim r^{-4}$. Here we define a non-equilibrium functional, $S_D$, which is maximized for DARKexp and increases monotonically during the evolution towards equilibrium of idealized collisionless systems of the Extended Spherical Infall Model. Systems that undergo more mixing more closely approach DARKexp. 
\end{abstract}


\section{Introduction}

It has been known for decades that elliptical galaxies, dark-matter halos, and N-body systems evolved in computer simulations appear to have universal, or nearly universal structures. In particular, the Navarro--Frenk--White \citep[NFW,][]{1997ApJ...490..493N} law describes the density profiles of equilibrium N-body systems grown in numerical simulations exceptionally well.
The NFW law is well established and has been thoroughly tested for the last twenty or so years, but despite this there is no consensus on its physical basis. There have been many attempts to explain the shape of the NFW density profile using different approaches, from theoretical to phenomenological \citep[e.g.,][]{sti87,pad02,man03,ma04,lu06,sal12,2013MNRAS.430..121P,bia16,ber19,2020GReGr..52...61W}. For example, \cite{pad02} used a theoretical argument to obtain the NFW density profile shape from the matter power spectrum, while \cite{ma04} used kinetic theory to investigate the phase-space distribution of dark matter particles. However, a full understanding of the evolution towards equilibrium still has not been reached.

From an astronomical point of view, this means that while we can reproduce them in computers, we do not fully understand the structures observed in the universe. From an academic point of view, the lack of such a theory is unsatisfactory since the fundamental principles behind statistical mechanics---the field that studies systems with very large numbers of particles---should apply to any physical system, regardless of the underlying forces between the particles.

Statistical mechanics is a fundamental physical theory of macroscopic systems, and provides a solid foundation for thermodynamics. Despite its sweeping success in most branches of physics, it is sometimes held that a statistical mechanical theory for many-body, purely collisionless systems dominated by gravity cannot be successfully formulated. In this study we outline such a theory{\footnote{The work presented here falls short of a full theory, which would ideally start with kinetic theory, and proceed to derive the entropy functional in a logically continuous manner. However, since our starting point is the same as Boltzmann's expression for the number microstates of a given macrostate, Eq.~\ref{eq:W} and following equations, we will refer to our work as `theory', keeping in mind the above caveat.} and demonstrate monotonic evolution towards a well-defined maximum entropy state. This equilibrium is furthermore exactly that seen in numerical simulations and in real astronomical systems.

\begin{figure*}
    \centering
    \vspace{-2.cm}
    \includegraphics[trim={0cm 5cm 0cm 2cm},clip,width=0.75\textwidth]{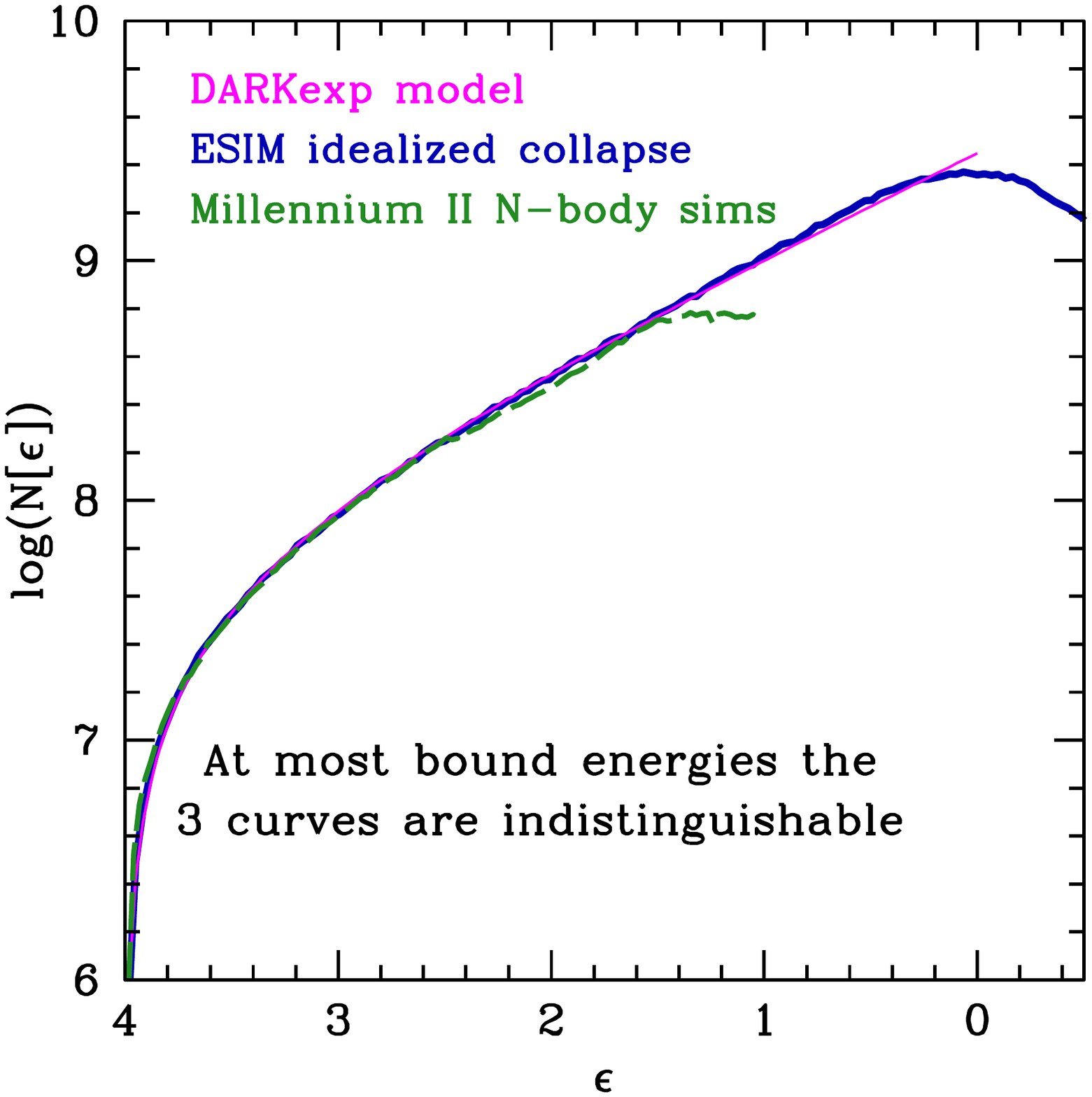}
    \caption{Energy distribution of our DARKexp model (magenta), and two numerical calculations: idealized ESIM described in this paper (blue), and Millennium II high resolution N-body simulation \citep[][green dashed]{boy09,2016JCAP...09..042N}.
    Three parameters were adjusted to fit the ESIM and N-body results to DARKexp: the depth of the central potential, $\Phi_0$, the inverse temperature $\beta$ (such that $\epsilon=\beta E$ is the dimensionless energy), and the amplitude. 
    The $n(\epsilon)$ {\it shapes} of N-body and ESIM systems are fitted very well by DARKexp. The fits are particularly good at the most bound, or most equilibrated energies: the three curves are indistinguishable.
    }
    \label{fig:DEM}
\end{figure*}

\section{Theory of the equilibria of collisionless self-gravitating systems}

To set the stage, we first review existing work on the equilibrium state. This will provide the background and motivation for the scope of this paper, namely to quantify the approach to equilibrium.

The challenges in formulating a statistical mechanical theory of self-gravitating systems include:
(1) The long-range, unshielded, nature of the Newtonian gravitational force, $F \sim r^{-2}$, which leads to a non-additive system \citep[e.g.,][]{cam14}, potentially of infinite extent.  On the other hand, in standard statistical mechanics the forces are short-range, and/or shielded, and therefore the systems are additive, where energies and entropies of sub-systems can be added together.
(2) The time-scale for the only well understood dynamical mechanism that can bring about relaxation, i.e., collisional relaxation, which proceeds through binary interactions, scales as $t_{\rm 2-body\,relax} \sim N(8\ln N)^{-1}Rv^{-1}$,
where $N$ is the number of particles, $R$ and $v$ are the system's characteristic size and internal velocity, and hence easily exceeds the age of the universe for most systems of interest. Collisionless relaxation, on other hand, proceeds on dynamical time-scale, which is shorter than the Hubble time.
(3) A successful theory must not only overcome these issues, but must also lead to predictions that are consistent with observed and simulated structures. A number of theoretical models can be found in the literature, but most do not fit simulated halos well.
(4) Finally, a complete theory should address the approach to equilibrium, ideally in the form of an $H$ theorem (2nd law of thermodynamics), or similar.
These challenges and their solutions are listed schematically in Table~\ref{tab:table1} and are briefly addressed below.
\begin{table*}
\caption{\label{tab:table1}%
Collisionless relaxation: challenges and solutions.
}
\begin{tabular}{lll}
&
\textrm{Challenge}&
\textrm{Solution}\\
\colrule
(1) & Gravity is a long range force & Incomplete relaxation; DARKexp (Paper I)\\
(2) & Time-scale of relaxation & $t_{\rm{violent\,relaxation}}\ll t_{\rm{2-body\,relaxation}}$ \citep{lyn67}\\
(3) & Agreement with simulations/observations & DARKexp (Paper II, III) \\
(4) & Approach to equilibrium & Non-equilibrium functional $S_D$ (this paper, Paper V) \\
\colrule
\end{tabular}
\end{table*}

The history of attempts to understand self-gravitating systems in terms of statistical mechanics is long and varied.
\citet{ogo57} (and later \citep{lyn67,shu78,mad87}) first formulated a particle statistical mechanical theory. Maximizing the number of microstates corresponding to a given macrostate, within the microcanonical ensemble \citep{1866Boltzmann} under constraints of conserved total mass and energy in six-dimensional phase space, he derived the isothermal sphere, $f(E) \sim \exp(-\beta E)$, where $f$ is the phase-space density (distribution function), 
$E$ is the energy per unit mass of a particle, and $\beta$ is an inverse
temperature. The resulting density structure, $\rho \sim r^{-2}$, has infinite mass, violating the constraint under which it was derived. This highlights the challenge related to the long-range forces involved in the problem (item 1 above; see Table~\ref{tab:table1}). Attempts to truncate these structures in terms of formation scenarios or incomplete relaxation have been partially successful, although they invariably involve aspects of arbitrariness and do not solve the issue of what are the structures of fully relaxed systems \citep{kin66,sti87,hjo91,hjo93}. Alternative approaches based on non-standard entropy functionals \citep{1999BrJPh..29....1T,1993PhLA..174..384P}
have not proven useful \citep{2008PhRvE..77b2106F,barnes07}.

\citet{lyn67} independently developed and refined the theory in two important respects. Pointing out the collisionless nature of the dynamics, as encoded in the collisionless Boltzmann (Vlasov) equation, he formulated the theory in terms of non-overlapping ‘phase elements’, leading to an exclusion principle and a resulting phase-space distribution akin to a Fermi--Dirac distribution, $f \sim (\exp(\beta (E-\mu))+1)^{-1}$. In the non-degenerate limit this is Ogorodnikov’s Maxwell--Boltzmann distribution. More importantly, Lynden-Bell highlighted the role of the efficient energy exchange between phase elements via a collective, rapidly varying gravitational potential, thus providing a physical mechanism for relaxation (addressing item 2 above) \cite[][see Fig.~1]{bar12}. In addition to this `violent relaxation', other mechanisms, for example, the radial orbit instability \cite[e.g.,][]{hus99,2006ApJ...653...43M,bar09} may help drive a stationary system towards equilibrium.

Just like the papers described above, most other attempts to apply statistical mechanics to gravity involve the distribution function, $f(E)$
\citep{2011A&A...526A.147K,2014PhRvD..90l3004B,2020GReGr..52...61W,1987MNRAS.229..103W,2013MNRAS.430..121P,2014PhR...535....1L}.
The resulting problems listed at the beginning of this section and summarized in Table~\ref{tab:table1} have led some to conclude that there is no maximum entropy solution for gravity. 

\citet[][Paper I of this series]{2010ApJ...722..851H} argued that while the distribution function is a central concept in many branches of physics, and relevant for self-gravitating systems dominated by collisions, in {\it collisionless} self-gravitating systems the quantity of physical significance is the differential energy distribution $n(E)=f(E)\,g(E)$, as the following paragraph explains. However, $g$, the density of states, and $f$, both of which are defined in the six-dimensional phase space, are not independent of each other. To obtain $f$ for a self-gravitating system one needs to know the system's Newtonian potential, $\Phi(r)$, i.e., its large scale structure. To obtain $g$ one also needs $\Phi(r)$. So for a self-consistent system, once $f$ is specified, one has no choice in what $g$ should be (this is unlike systems of electrons, etc.).

The key ansatz in Paper I is that the steady-state configuration of a collection of collisionless self-gravitating particles has to be statistically the most likely state. This is the principle of maximum entropy. The derivation, outlined in the next section, follows closely that of the Maxwell--Boltzmann distribution, with the key difference of using a better approximation for natural logarithms of factorials than the Stirling approximation, and interpreting the resulting occupation number as the differential energy distribution, $n(E)$, instead of the distribution function, $f$. We choose the state space to be the one-dimensional energy space, rather than the six-dimensional phase space, because particles in collisionless equilibrium stay on fixed orbits with constant energy 
\footnote{This corresponds to assigning non-equal {\em a priori} weights to
six-dimensional phase space cells \citep[][Paper I]{sti87}, reflecting the fact that not all phase space cells are equally accessible.},%
\footnote{For systems with non-isotropic velocity distribution, the state space would have included angular momentum in addition to energy \citep[][Paper IV]{2014ApJ...783...13W}}. With these modifications, the equilibrium solution, known as DARKexp, has the form,
\begin{equation}
n_{\rm D}(E)\propto e^{-\beta(E-\Phi_0)}-1,\label{nD}
\end{equation}
where $\Phi_0$ is the Newtonian potential at the center of the system, $\Phi_0 \le E \le 0$, and $\beta < 0$, i.e., a negative `temperature' (which should not be mistaken for the kinetic temperature \footnote{However, for collisional  systems,  such as the isothermal sphere, the kinetic temperature is the same as $1/\beta$ in the distribution function.}). The particle energy per unit mass, $E$, is the sum of kinetic and potential energies. The fact that we are dealing with gravity comes in through the latter. Equation~\ref{nD} provides an excellent description of the energy distribution, density profiles and velocity dispersion profiles of simulated dark-matter halos \citep[][the last two are Papers III and IV of this series]{2015ApJ...811....2H,2016JCAP...09..042N,2018JCAP...02..033Y,2010ApJ...725..282W,2014ApJ...783...13W}, and the density profiles of equilibrium observed galaxy clusters \citep{2013MNRAS.436.2616B,2016ApJ...821..116U}.

Figure~\ref{fig:DEM} shows three energy distributions: the DARKexp model (magenta), an average of many dark matter halos from high resolution Millennium II N-body simulations \citep[green dashed;][]{boy09,2016JCAP...09..042N}, and an average of many halos of a simplified collapse model (blue), discussed in \cite[][which is Paper II of this series]{2010ApJ...722..856W}, and later in this paper. Though these three $n(E)$'s have different origins---statistical mechanics theory, fully non-linear numerical simulation, and an idealized numerical collapse---they all represent the final equilibrium state of self-gravitating systems. 

All this largely settles item 3 above, but leaves item 4 unanswered. This 
paper addresses the latter challenge, namely the approach to equilibrium.

\section{A new `entropy'-like functional}

The remarkable agreement between the three distributions at bound particle energies, $\epsilon\gtrsim 1$ (Figure~\ref{fig:DEM}) supports our assertion that DARKexp is the correct description of collisionlessly relaxed systems, and provides motivation for extending the theory, namely, seeking a functional related to DARKexp that will increase during evolution and reach an extremum at equilibrium. If successful, such a theory will represent gravity's analogue of the 2nd law of thermodynamics. 

\begin{figure*}
    \centering
    \vspace{-1.8cm}
    \includegraphics[trim={0cm 5cm 0cm 0cm},clip,width=0.49\textwidth]{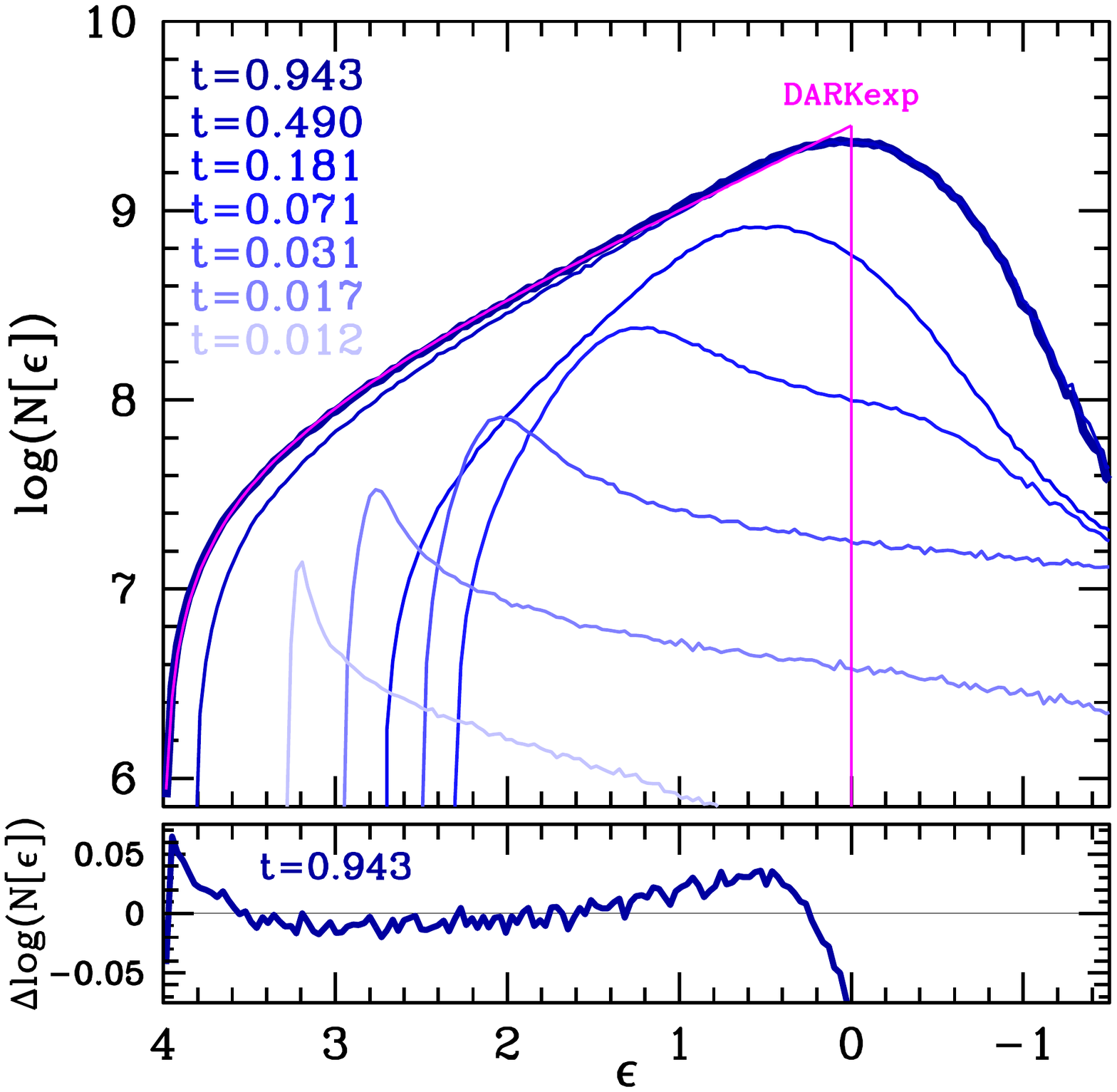}
    \includegraphics[trim={0cm 5cm 0cm 0cm},clip,width=0.49\textwidth]{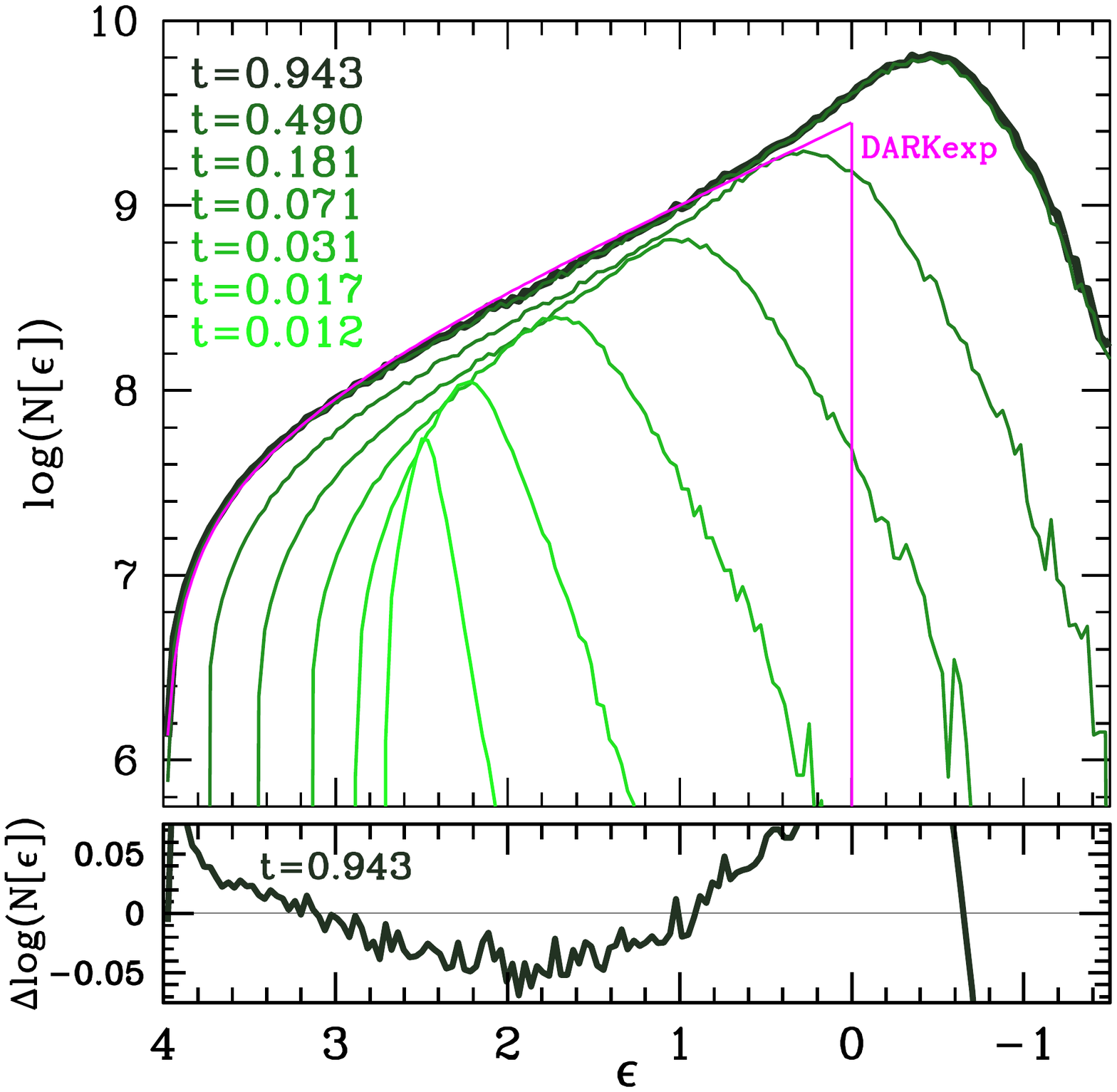}
    \vspace{-0.5cm}
    \caption{Evolution of the differential energy distribution, $n(\epsilon)$, of two ESIM halos (called A and E in later figures) for 7 cosmic epochs labelled by time, in units of the present epoch. Lighter shades represent earlier epochs; the latest epoch is represented by thick lines. The number of shells was not normalized to the final epoch, so the earlier epochs have smaller number of collapsed shells and hence lower normalizations. (All later figures show these two systems in blue and green, respectively.) The zero-point of the dimensionless energy $\epsilon=\beta E$, was set to approximately where the energy distribution turns over, on the right side. 
    The magenta curve is a DARKexp fit to the final state, with $\beta\Phi_0=\phi_0=4$. The residuals of the final state are shown in the bottom panel.}
    \label{fig:NE}
\end{figure*}

Addressing the evolution of collisionless systems, Tremaine, Henon \& Lynden-Bell \citep{1986MNRAS.219..285T} demonstrated that coarse graining and phase mixing would lead to $H(t_2) \ge H(t_1)$ for $H=-\int C(F) d\tau$, where $C$ is a convex function and $F$ is the coarse-grained distribution function. However, $dH/dt \ge 0$ could not be demonstrated without introducing additional assumptions \citep{1996ApJ...457..287S,1987MNRAS.225..995K,1987JApA....8..257S,1987ApJ...320..477D,2005MNRAS.360..892D}. 
Spergel \& Hernquist \citep{1992ApJ...397L..75S} proposed an ansatz for a collision term, which would lead to monotonic evolution, but did not demonstrate that the evolution of N-body systems lead to monotonic evolution of an $H$ function.

Our strategy is to use the derivation of DARKexp itself, i.e., the maximum entropy state, to come up with the form of the functional, and then test if it changes monotonically as the system evolves.
From a statistical point of view, entropy is related to the number of microstates associated with a given macrostate of a system. In a collisionless Newtonian system, any macrostate is fully described by its differential energy distribution, $n(E)$. The corresponding number of microstates is the number of possible ways of distributing $N$ particles among $j$ energy levels, each with degeneracy (cell size) $g_j$. Since it is a basic assumption of DARKexp that all energy levels have equal {\it a priori} probabilities, we set $g_j$ to be a constant, $g$. Hence, $W=N!\,\Pi_j({{g}^{n_j}})/({n_j!})$. Using the Gamma function as the continuous limit of the factorial, we write,
$\ln W=\ln N!+\sum_j n_j \ln g-\sum_j\ln\Gamma(n_j+1).$

The final equilibrium state of a system is the one that has the maximum possible microstates associated with it, subject to the constraint of total fixed energy, which in the final state is proportional to $\sum_j n_j E_j=E_t$\footnote{The actual total energy, $E_{\rm tot}$, which is the sum of kinetic and potential terms of an equilibrium system is $\sum n_j E_j = 3E_{\rm tot}$. The factor of $3$ arises due to the binding energy.}. The constraint is introduced through a Lagrange multiplier term, $-\beta E_t$, added to the expression for $\ln W$.

In the classical derivation one would also constrain the equilibrium system to have a fixed total particle number.\footnote{This was done in Paper I, and led to exactly the same DARKexp form, indicating that constraining total $N$ is not essential to the derivation. (For completeness, that derivation is reproduced in this paper's Appendix~\ref{sec:app}).}
However, since our goal is to obtain the {\it shape} of the equilibrium $n(E)$, not its normalization, there is no need to use a Lagrange multiplier for $N$. The only situation when one would pay attention to $N$ is when it is infinite. Fortunately, that is already precluded by the use of $\beta$ Lagrange multiplier term, which is proportional to $\sum_j n_j E_j$. 

Thus, we define a new functional as $\ln W-\beta\sum_j n_j E_j$, but leaving out the terms that contain $N$,
\begin{eqnarray}
S_D&=S_\Gamma - \beta E_t= \hspace{5cm}\nonumber\\
&=-\int \ln\Bigl\{\Gamma[n(E)+1]\Bigr\} dE-\beta\int n(E) E dE.
\label{eq:SD}
\end{eqnarray}

Before proceeding, we briefly show that extremizing $S_D$ yields DARKexp, and that the extremum is, in fact, a maximum. 
Varying $S_D$ with respect to $n$ leads to an extremum of 
$\delta S_D = - \psi(n+1)-\beta E$ = 0 and 
$\delta^2 S_D = - d \psi(n+1)/dn$,
where $\psi(n)=d \ln \Gamma / d n$ is the digamma function.
Using the approximation $\psi(n+1) \approx \ln(n+\zeta)$ 
with $\zeta \approx 0.561459\dots$
(Paper I) leads to equation (1), and 
$\delta^2 S_D = - (n+\zeta)^{-1} < 0$, i.e., 
{\it DARKexp is the maximum `entropy' state related to $S_D$.}
The act of finding the most likely state isolates DARKexp as the unique family of solutions among the infinity of solutions admitted by the collisionless Boltzmann equation.

We note that $S_D$ of Eq.~(\ref{eq:SD}) bears a resemblance to Massieu free entropy function \citep{massieu1869,callen1985}. However, in standard thermodynamics, $\beta$ is proportional to the inverse temperature. In our case, $\beta$ is not the inverse of the kinetic temperature of the system; in fact, there is no simple relation between the two. Furthermore, there is no single kinetic temperature that describes the entire collisionless self-gravitating system, even in equilibrium. The temperature (i.e., the velocity dispersion) varies with position within the system.

\section{The Extended Secondary Infall Model (ESIM)}

In this section we show that in an idealized self-gravitating collapsing system $S_D$ does increase as evolution progresses, and attains a maximum at equilibrium, whose $n(E)$ shape is very closely described by DARKexp.  Thus $S_D$ fulfils its intended purpose.

We use a simplified numerical model of gravitational collapse and relaxation of an isolated system, the Extended Secondary Infall Model (ESIM). In short, the halos consist of spherically symmetric shells, and have 3 phase-space dimensions: radial spatial, radial velocity and tangential velocity. The collapse proceeds from inside out, and leads to shell-crossing, i.e., shell overlap. Halos reach stable equilibrium at the present epoch.

ESIM \citep{ryd87,wil04} is different from commonly used N-body simulations. An N-body system consists of many ($\sim 10^4-10^9$) particles of the same mass, which move in 3 spatial dimensions under the force of Newtonian gravity. Each particle can explore the full 6D phase space, similar to what dark matter particles do in a real halo, as it evolves towards an equilibrium steady state. The main drawbacks of N-body simulations are (i) the high degree of complexity of the resulting dynamics, and (ii) the poor mass and force resolution in the densest central regions of halos.

\begin{figure}
\vspace{-1cm}
\includegraphics[trim={0cm 5cm 0cm 0cm},clip,width=0.48\textwidth]{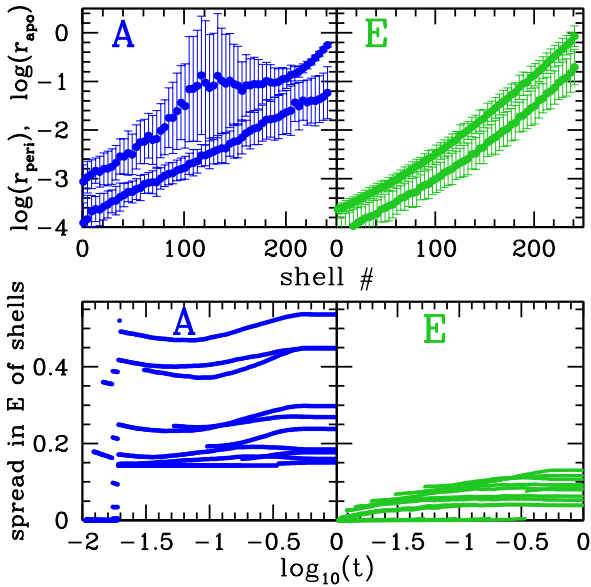}
\vspace{-0.5cm}
\caption{Comparison of the degree of mixing in two halos shown in Figure~\ref{fig:NE}, A ({\it left}), and E ({\it right}). {\it Top:} Peri-centers (lower band) and apo-centers (upper band) of ESIM shells, at the present epoch, in units of the virial radius. For both peri- and apo-centers we show the average values as filled dots, and the rms dispersion as error-bars. 
{\it Bottom:} Rms dispersion in energies of 10 shells, as a function of cosmic epoch. For any given shell and epoch, the rms is over many individual realizations of the same halo. To facilitate comparison, these rms values have been normalized by the average energy of that shell. Halo A shows larger dispersion in shell's energies (larger rms), compared to halo E, implying that A can undergo more mixing during evolution.}
\label{fig:mixing}
\end{figure}

ESIM systems give up some degree of realism, but gain in these two aspects. ESIM halos are geometrically spherically symmetric, and have only 3 phase-space dimensions: 1 radial and 2 velocity, radial and tangential. 

Instead of particles of constant mass, they have shells whose masses decrease towards the central densest regions, leading to higher numerical resolution where it is most needed. For example, in the ESIM halo in Figure~\ref{fig:DEM} the number of shells per energy bin at the most bound energies is about a factor 3--4 larger than at near-escape energies, which means that individual shells at most bound energies are about 1000 times less massive than those at near-escape energies. Before collapse all shells are infinitely thin, with each placed at its radius of maximum expansion.  In addition to this radial structure, initial conditions also include angular momentum for each shell, $J_i$, specified at some early cosmic epoch. Before assigning $J_i$ to shells one needs to specify the average $\bar J_i(r)$, as a function of radius within the initial halo, as well as the shape of the distribution around that average, PDF($\Delta J_i=J_i-\bar J_i(r)$), and its width. The width of this distribution will play an important role in halo evolution. Each shells' $J_i$'s are picked randomly from this distribution. Each halo presented in this paper is an average of many realizations (few hundred), all sharing the same initial density profile, $\bar J_i(r)$, and PDF($\Delta J_i$). We have tried several types of $\bar J_i(r)$ and PDF($\Delta J_i$) to ensure that our conclusions hold beyond the cases presented here.

Calculation of halo evolution proceeds as follows. Halos collapse from inside out, one shell at a time. As the inner shells collapse, the outer shells are fixed at their initial radii. This means that even though there is no Hubble expansion as such (outer shells are not expanding), ESIM implements the retardation of collapse due to cosmological expansion, because shells collapse sequentially, and not all at once. 

As each shell falls in, it acquires a finite radial extent, with minimum and maximum radii, $r_{\rm peri}$ and $r_{\rm apo}$. Thus the mass of each shell is spread over a range of radii; the amount of mass it contributes at a given radius depends on how much time it spends there, which is $\propto\!{v_r}^{-1}$, where $v_r(r)$ is its radial velocity.  The typical ratio $r_{\rm apo}/r_{\rm peri}\!\sim\!{\rm few}\!-\!100$, so there is considerable overlap between shells, also known as shell-crossing. In any given system, the width of the PDF($\Delta J_i$) roughly determines the actual range of $r_{\rm apo}/r_{\rm peri}$, and hence the amount of shell overlap, which in turn determines the degree of mixing that a system experiences (see Section~\ref{sec:evol}). Here, mixing refers to the energy exchange between individual shells and the global potential, and facilitates relaxation.

Each shell conserves its radial action and angular momentum at each step in the evolution, and satisfies the specific energy equation: $E_i(t)=\Phi(r,t)+\frac{1}{2}\Big[(J_i/r)^2+v_r^2(r,t)\Big].$ (This is different from N-body halos, where particles' actions can change, to some degree.) The shells' peri- and apo-centers, $r_{\rm peri}$ and $r_{\rm apo}$, are radii beyond which $v_r^2$ becomes negative. As successive shells fall in, the energies, peri- and apo-center radii of all the interior shells are recalculated, to agree with the changing potential, $\Phi(r,t)$. As the collapse proceeds, the inner shells settle to a nearly steady equilibrium configuration because the potential at small radii ceases to change appreciably, while still continuing to vary at larger radii. 
ESIM evolution does just one thing over and over again: it recalculates the energies and radii of each shell to agree with the changing potential. Changing energies of individual particles/shells per unit mass is an essential aspect of collisionless relaxation \citep{lyn67,pad02}.

\begin{figure}
    \centering
    \vspace{-1.75cm}\hspace*{-1.0cm}
     \includegraphics[trim={0cm 5cm 0cm 0cm},clip,width=0.49\textwidth]{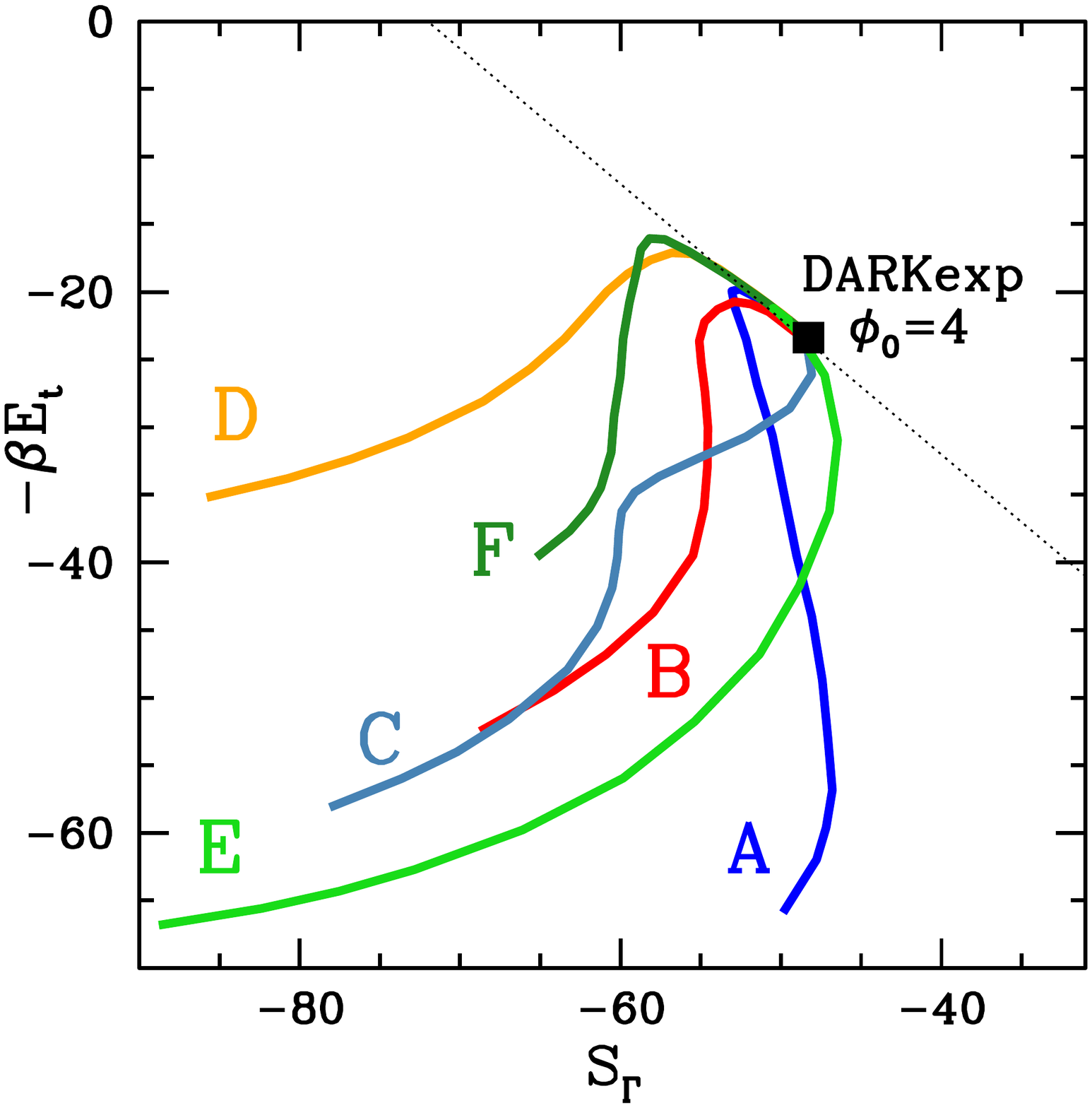}
     \vspace{-0.25cm}
    \caption{Evolution of the two individual terms of Eq.~(\ref{eq:SD}) in the plane defined by them. The 6 ESIM halos are labelled A--F. Early epochs are at the lower left, while the final epoch is in the upper right, and is nearly coincident for all 6 halos. The coordinates of the black square were obtained independently of the color tracks, by plugging the equation of DARKexp $n(E)$, with $\phi_0=4$, into the two terms of Eq.~(\ref{eq:SD}), using the same normalization of $n(E)$ as used for the 6 colored tracks. The dotted black line going through the black square has constant $S_D$.}
    \label{fig:2Devol}
\end{figure}

The use of the energy equation means that individual shells do not interact with each other directly; no forces between individual shells are calculated. Instead, shells adjust themselves to the global potential, making ESIM a mean-field calculation. This also ensures that ESIM dynamics is collisionless.

To sum up, ESIM is an idealized version of gravitational collapse. It strips away a number of dynamical processes operating in full N-body simulations, like, radial orbit instability, making ESIM evolution differ from that of N-body simulations. But even though the pathway to equilibrium may be different in ESIM compared to simulation, both types of dynamics drive the system to equilibrium, and both arrive at the same final state represented by DARKexp, as shown in Figure~\ref{fig:DEM}.

\section{Evolution towards equilibrium}\label{sec:evol}

We evolved a large number of ESIM halos, with a variety of initial conditions; 
we present 6 of them here, labeled A--F. 

Figure~\ref{fig:NE} shows the evolution of energy distribution of two example halos, A and E, whose initial distributions of angular momenta for each shell (PDF[$\Delta J_i$]) were a wide top-hat, and a Maxwellian, respectively.  To fit DARKexp model to $n(E)$ we adjusted three parameters: the depth of the central potential, $\Phi_0$, the inverse temperature $\beta$, and the amplitude. The last parameter, which is related to the total number of shells, has no physical significance. The parameter $\beta$ makes all energies unitless, $\epsilon=\beta E$, and $\phi_0=\beta \Phi_0$, where $\phi_0$ controls the extent of the exponential portion of DARKexp.

These three parameters were fitted to the final equilibrium halos, which are shown as thick dark blue and dark green for halos A and E, respectively. The values of $\beta$ and $\phi_0$ were then applied to the earlier epochs, resulting in $n(\epsilon)$ distributions shown in thin lighter blue and green colors. DARKexp of $\phi_0=4$ (magenta curve) provides a remarkable representation of the final state at bound energies, where mixing and relaxation were most effective.

\begin{figure}
    \centering
    \vspace{-2.7cm}\hspace*{-2.2cm}
    \includegraphics[trim={0cm 5cm 0cm 0cm},clip,width=0.7\textwidth]{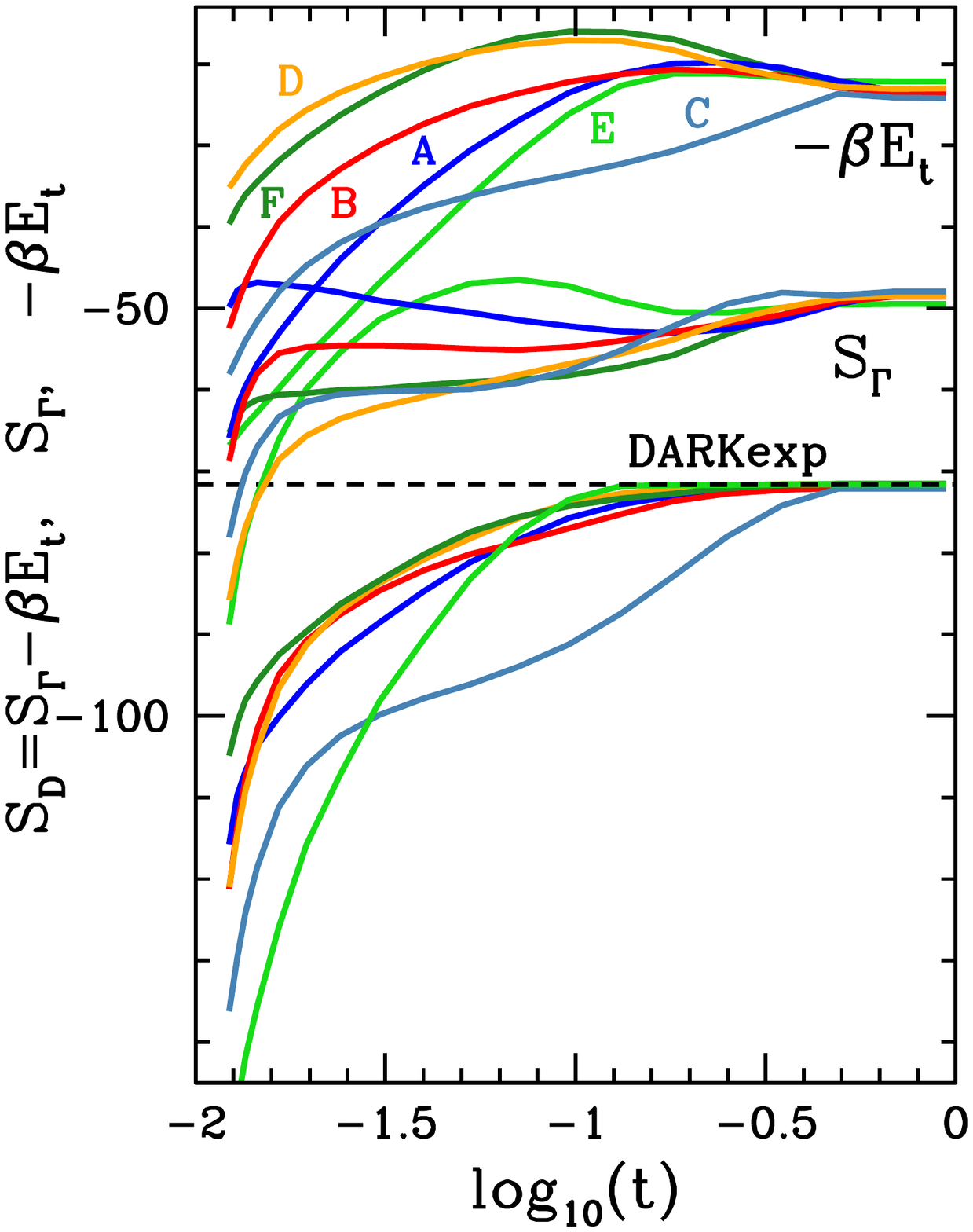}
    \vspace{-0.5cm}
    \caption{Evolution of the proposed functional, $S_D=S_\Gamma-\beta E_t$, Eq.~(\ref{eq:SD}), as well as $S_\Gamma$, and $-\beta E_t$ vs.\ log of cosmic time. The 6 ESIM halos are labelled A--F, as in Figure~\ref{fig:2Devol}. $S_\Gamma$ and $-\beta E_t$ approximately mirror each other's behavior for most of evolution. While $S_\Gamma$ increases monotonically only at late times (when the systems are close to equilibrium), $S_D$ increases monotonically throughout the evolution. If a system has $S_D$ value corresponding to that of DARKexp, the shape of its energy distributions can only be that of DARKexp. All systems evolve to DARKexp of $\phi_0=4$, represented by the black dashed line.  This is the key result of our paper. 
    }
    \label{fig:SESG}
\end{figure}

Halos A--F were given different shapes and widths of PDF($\Delta J_i$), which resulted in different amount of mixing during evolution. Figure~\ref{fig:mixing} illustrates halos A and E, using two measures of mixing. The top panels show the average and rms dispersion of apo- and peri-centers of shells at the final epoch. Each shell is represented by two solid dots: the top (bottom) is the average apo-center (peri-center) of 300 individual realizations that make up a halo, and the error-bars show the rms dispersion in these. The fact that shells show considerable overlap is evidenced by the fact that a horizontal line at any fixed $r$ will intersect many shells, i.e., many shells pass through any given radius. The width of the rms distribution illustrates another aspect of halo evolution: a large rms means that different realizations of the same system---specified by initial density profile, $\bar J_i(r)$, and PDF[$\Delta J_i$]---chose very different evolutionary paths. In other words, systems with larger rms (illustrated here by halo A) explored a larger fraction of the available energy space, i.e., underwent more mixing in energy space than systems with smaller rms (halo E).

The bottom panels show the dispersion in the shells' energies for 10 shells as a function of epoch, normalized by the average energy for that shell. As for the peri- and apo-centers, the average and the rms are over the many realizations of a given system. The energy dispersion is significantly higher in halo A compared to E, again indicating that halo A is more mixed that E. The two measures of mixing correlate with each other: halo A is better mixed than E.  Halo A is better modeled by DARKexp, suggesting that more mixing during evolution leads to closer adherence to the maximum of functional $S_D$. Among the many halo initial conditions we tried there were cases where the rms dispersion in shells' peri- and apo-centers, and energies were smaller than in halo E; these halos deviated from the DARKexp shape at the present epoch.

Having examined the evolution of the halos' $n(E)$ and mixing in energy space, we now turn to the main topic of this paper, functional $S_D$, Eq.~(\ref{eq:SD}). Figure~\ref{fig:2Devol} shows how the same 6 halos evolve in the plane of $S_\Gamma$ vs. $-\beta E_t$, the two terms of $S_D$. They take different paths, but end up at the same location, which is nearly coincident with DARKexp of $\phi_0=4$, marked by the black square. The coordinates of the black square were obtained independently of the 6 color tracks, by plugging the analytical equation of DARKexp $n(E)$, with $\phi_0=4$, into the two terms of Eq.~(\ref{eq:SD}), using the same normalization of $n(E)$ as used for the 6 halo tracks. The dotted black line going through the black square has constant $S_D$, and the black square represents the extremum of $S_D$. It is apparent from the plot that the end point of evolution of ESIM halos is DARKexp, and that it is a maximum of $S_D$.

Figure~\ref{fig:SESG} presents the evolution of the two terms in $S_D$, Eq.~(\ref{eq:SD}), as a function of cosmic time, for the 6 halos. All 6 tracks are different, but the end points of all 6 are at or very near DARKexp, represented by the dashed line. We note that halos with much smaller degree of mixing than A--F (not shown here) do not end up near DARKexp in their final epoch, further suggesting the importance of mixing in attaining the DARKexp shape corresponding to the maximum of functional $S_D$.
While the evolution of the two terms in Eq.~(\ref{eq:SD}) is not monotonic (upper part of Figure~\ref{fig:SESG}), the evolution of the functional $S_D$, as a function of time, is monotonic, as shown in the lower part of Figure~\ref{fig:SESG}. This demonstrates an apparent 2nd law of thermodynamics.

The density and velocity anisotropy profiles of all 6 halos are shown in Figure~\ref{fig:density}. These are very similar to profiles of N-body simulated halos. The DARKexp $n(\epsilon)$ parameter $\phi_0$, which quantifies the dimensionless potential depth, results in different density profile shapes at small radii. As such, DARKexp has one free shape parameter, $\phi_0$ \citep{2015ApJ...811....2H}.

\begin{figure}
    \centering
    \vspace{-1.75cm}\hspace*{-2.3cm}
     \includegraphics[trim={0cm 5cm 0cm 0cm},clip,width=0.7\textwidth]{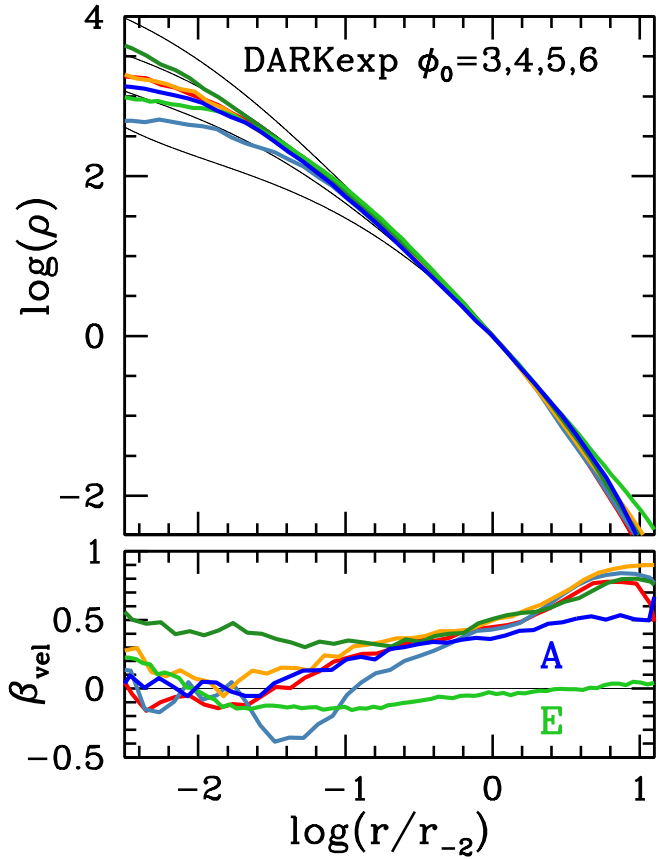}
     \vspace{-0.5cm}
    \caption{{\it {Top:}} Density profiles of the 6 ESIM halos A--F, represented with the same colors as in Figure~\ref{fig:SESG}. Black thin curves show DARKexp of $\phi_0=3,4,5$ and $6$ (least to most cuspy at center). The distance from the halo center, $r$, is in units of $r_{-2}$, where $d\ln\rho/d\ln r=-2$. At very small radii the profiles are somewhat affected by numerical artifacts. {\it{Bottom:}} Profiles of velocity anisotropy, defined as $\beta_{\rm vel}=1-(\sigma_\theta/\sigma_r)^2$, where $\sigma_\theta$ and $\sigma_r$ are velocity dispersions in the radial and tangential directions, respectively.}
    \label{fig:density}
\end{figure}

\vspace{0.5cm}
\section{Future work}

What we have presented here and in the previous papers of the same series is a statistical mechanical approach to relaxation and equilibrium state of perfectly collisionless self-gravitating systems. As already stated above, our work does not form a complete theory of the formation of dark matter halos. For a comprehensive theory one needs a kinetic theory approach, \citep[e.g.,][]{dev02,ma04,bia16,cha21,bar22}, as well as a solid connection to thermodynamic concepts \citep[e.g.,][]{bou10}. However, no such theory exists at present for virialized collisionless systems. On larger, but non-linear scales much theoretical progress has been made to understand and accurately account for the matter power spectrum \citep[e.g.,][]{bar19}. However, the published attempts to reproduce the structure of virialized dark matter halos based on kinetic theory do not make a connection to observations or simulations by presenting, for example, a comparison between their predictions of the density profiles and velocity structure of dark matter halos, especially of the central densest regions, and the results of high resolution N-body simulations.

A future full theory could take inspiration from recent developments in a different, though related field of collisional relaxation, where particle-particle encounters are important. In astrophysical context the relevant systems are self-gravitating thin and thick stellar disks, as well as globular and nuclear star clusters \citep[e.g.,][]{rau96,hey10,fou18}.

\vspace{0.5cm}
\section{Conclusion and Discussion}

The DARKexp energy distribution, $n(E)$ was obtained by extremizing the expression for the number of microstates subject to constraints of total energy. We have shown that this extremum is, in fact, a maximum. 
Because the sum of $\ln W$ and the energy Lagrange multiplier term attains a maximum in the steady state, we put forth the ansatz that the sum should steadily increase during the evolution. Ignoring constant terms in that expression, leaves two terms, which constitute our new functional, $S_D$, Eq.~(\ref{eq:SD}).  Using an idealized gravitational collapse model, ESIM, we show that the value of $S_D$ does increase monotonically and approaches that of DARKexp at equilibrium, for a wide range of halo initial conditions. The shape of $n(E)$ also approaches that of DARKexp, becoming almost indistinguishable from it at the most bound energies. This is only expected in truly collisionless systems, and requires a sufficient amount of mixing during evolution. 

DARKexp makes two predictions: (1)
One of the features of DARKexp is that it should apply to both cold and warm dark matter halos, because the derivation does not make a distinction between the two. This is, in fact, borne out in simulations; \cite{1999MNRAS.308.1011H} show that universal density profiles emerge regardless of cosmology and the warm/cold nature of dark matter. (2) The parameter $\phi_0$ in the expression for $n(E)$ is a variable in our theory. When converted to density, it affects the inner density profile slope, allowing it to range from flat cores to nearly `isothermal', $\rho\propto r^{-2}$ cusps, while keeping the outer structure unchanged. So DARKexp predicts a diversity in inner density slopes, but since it is a statistical mechanics theory, it cannot anticipate what astrophysical drivers will realize that prediction.  In fact, a range of inner slopes has been observed in dark matter dominated dwarf galaxies, and in N-body simulations. In simulations, the slope has been shown to correlate with the galaxies' merger history \citep{2010MNRAS.402...21N,2015ApJ...811....2H,lud13,die22}.

The physical driver of evolution is the fluctuating Newtonian potential, which allows a halo to visit many microstates during evolution. Taken collectively, ESIM shells explore the one dimensional energy space reasonably well, even though the $N$-dimensional space of energies of $N$ shells is very sparsely covered. Nonetheless, ESIM evolution allows the final state to be very close to the analytically derived DARKexp $n(E)$ shape, for all practical purposes.

This paper presents the first demonstration of a functional that increases monotonically with time during evolution, attains a maximum at equilibrium, and predicts the shape of the equilibrium energy distribution. In this framework the approach to equilibrium and the equilibrium state itself form a self-consistent theory.

Our proposed functional $S_D$ should also describe the evolution of N-body simulated systems, just like DARKexp accurately describes the final state of these systems. We expect that the changes in $S_D$ may not be completely monotonic with time, but will experience small bumps due to halo mergers and uneven accretion of matter. It will be interesting to test this in the future. 

If the main conclusion of this paper holds for N-body and real systems, namely, that the new functional $S_D$ is maximized during evolution, it would explain the (near) universality of their equilibrium states.

\vspace{0.5cm}\noindent
\begin{acknowledgments}
We would like to thank the anonymous referee for their insightful comments. We thank Scott Tremaine for challenging us to demonstrate the existence of an entropy-like functional that evolves monotonically with time and reaches DARKexp as the maximum entropy state. We are grateful to Jeppe Dyre, Rafael Fernandez, Alex Kamenev, Yongzhong Qian, Panagiotis Tolias, Jenny Wagner and Radek Wojtak for insightful comments. JH was supported by a VILLUM FONDEN Investigator grant (project number 16599).
\end{acknowledgments} 

\bibliographystyle{aasjournal}

\providecommand{\noopsort}[1]{}\providecommand{\singleletter}[1]{#1}%

\appendix

\section{The derivation of DARKexp}

This derivation was originally presented in \cite{2010ApJ...722..851H}. We reproduce it here for completeness.\label{sec:app}

Our goal is to derive the most likely equilibrium state of a collection of self-gravitating classical particles that do not obey exclusion principle.  We use the tools of standard statistical mechanics \citep{bol1896} to calculate the state (macrostate) with the maximum number of associated microstates. The number of microstates is given by
\begin{equation}
  W=N!\,\prod_{i}\,\frac{g_i^{n_i}}{n_i!},\label{eq:W}
\end{equation}
where $n_i$ is the number of particles in an energy level $i$, $g_i$ is the degeneracy of that energy level, and $N=\sum_{i}\,n_i$ is the total number of particles in the system. Before maximizing $W$, we convert Eq.~(\ref{eq:W})  to natural logs, and recall that $\Gamma(n+1)=n!$. We want to find the state with the maximum $W$ under the constraint of fixed particle number and fixed total energy, so we add two Lagrange multiplier terms, one for each of these imposed conditions. The new expression becomes,
\begin{equation}
  \ln W=\ln N! + \sum_i\,n_i\,\ln g_i - \sum_i\,\ln\Gamma(n_i+1)
  +\alpha(N-\sum_i\,n_i) + \beta(E_t-\sum_i\,n_i E_i),\label{eq:lnW}
\end{equation}
where $E_t$ is fixed total energy, and $E_i$ is the energy of level $i$. Note that while $\alpha$ is unitless, $\beta$ has units of inverse energy. Extremizing Eq.~{\ref{eq:lnW}} we get,
\begin{equation}
  \frac{d\ln W}{d\ln n_i}=0+\ln g_i - \frac{d\ln\Gamma(n_i+1)}{dn_i}-\alpha-\beta E_i=0.\label{eq:dlnW} 
\end{equation}
Because the distribution of particles in a real system is not expected to be step-like, we switch from a discrete to continuous limit of the Gamma function, $\Gamma(n)=\int_0^\infty\,t^{n-1}\,e^{-t}\,dt$. Then, $d\ln\Gamma(n)/dn=\psi(n)$, which is the digamma function. For large values of $n$, $\psi(n+1)\approx \ln n$, which is the Stirling approximation. In our case, small occupation numbers $n_i$ need to be considered, so instead of the Stirling approximation we use a more accurate one, $\psi(n+1)\approx\ln(n+\zeta)$, where $\zeta=e^{-\gamma}$, and $\gamma=0.577215...$ is the Euler's constant.  Such approximations have been suggested as early as the 1950's \citep{lan54}. With this Eq.~(\ref{eq:dlnW}) reads,
\begin{equation}
  \ln g -\ln(n+\zeta)-\alpha-\beta E=0.
\end{equation}
Exponentiating this expression and rearranging the terms we get,
\begin{equation}
  n=g\,\exp(-\beta[\frac{\alpha}{\beta}+E])-\zeta=g\,\exp(-\beta[-E_0+E])-\zeta.
\end{equation}
To simplify this expression and determine the value of $g$ and the meaning of $E_0$, we note that particles in a self-gravitating system cannot have an energy lower (i.e., more bound) that a certain minimum, which corresponds to particles sitting at the bottom of the potential well, and having zero kinetic energy. If that energy is $E_0$ then the corresponding occupation number $n$ is zero: $0=g\exp(0)-\zeta$, therefore $g=\zeta$. The final expression for the occupation number becomes,
\begin{equation}
  n=\zeta\Big[\exp(-\beta[E-E_0])-1\Big]\propto \Big[\exp(-\beta[E-E_0])-1\Big].\label{eq:darkexp}
\end{equation}
The last expression follows because the total number of particles, or the total mass of the system is irrelevant.

The most important aspect of the derivation of DARKexp, Eq.~(\ref{eq:darkexp}), is that we interpret the occupation number $n$ not as the distribution function, $f(E)$, but as the (differential) distribution of particles, or mass, in energy, $dM/dE$. For a purely collisionless, isotropic system in equilibrium the energy of any given particle is constant. We use this fact to motivate our partitioning of the state space in cells of constant energy.

\end{document}